\documentclass[12pt,epsf,epsfig,wrapfig]{article}
\usepackage{graphicx}
\begin{document}
\begin{center}{\bfseries THE NUCLEON SPIN STRUCTURE AND QCD SPIN PHYSICS \footnote{Opening Lecture at the Workshop DUBNA-SPIN 07, Sept. 02-09, 2007, Dubna (Russia).}
\vskip 4mm
Jacques Soffer 
\vskip 4mm
{\small {\it Department of Physics, Temple University,\\
Philadelphia, Pennsylvania 19122-6082, USA\\
E-mail: jsoffer@temple.edu}}}
\end{center}
\vskip 4mm
\begin{abstract}Our knowledge of the nucleon spin structure has greatly improved over the last twenty years or so, but still many fundamental questions remain unsolved. I will try to review some of the puzzling aspects of the structure of the nucleon spin, in particular, what is known, what remains to be discovered and the prospects for the near future. I will also focus on some current activities in QCD spin physics. 
\end{abstract}
\vskip 5mm 
\section{Introduction}
Among the essential goals of QCD spin physics one has first, to understand the nucleon spin structure in terms of its basic partonic constituents and second, to test the {\it SPIN SECTOR} of perturbative QCD, at the highest possible precision level. Concerning the first point, one needs to know how the quark and gluon distributions in a polarized nucleon make its spin one-half and several questions arise in particular: what is the role of the orbital angular momentum? The second point is very relevant to reinforce the validity of the already well established perturbative QCD theory, because many spin asymmetries have been calculated, at the
next-to-leading order (NLO), and have not yet been compared with experimental data. Therefore
it is very legitimate to ask to what extent they will agree. 
We will try to answer the following questions:
What is known? What is missing? What needs to be measured next?
What are the prospects?\\
The basic information comes from Deep Inelastic Scattering (DIS), unpolarized
${\it l} N \to {\it l}^{'} X $, or polarized $\overrightarrow{\it l} \overrightarrow N \to {\it l}^{'} X$.
In the unpolarized case, widely measured over the last three decades, one gets access to $F_2^{p,n}(x,Q^2)=\sum_q e_q^2[x q(x,Q^2) + x\bar q(x,Q^2)]$. Here 
the $q(x,Q^2)$'s (same for antiquarks) are defined as $q=q_+ + q_-$, where $q_{\pm}$ are the quark distributions in a polarized
proton with helicity parallel $(+)$ or antiparallel $(-)$ to that of the proton.
In the polarized case, one measures the corresponding polarized structure function, $g_1^{p,n}(x,Q^2)=1/2\sum_q e_q^2[\Delta q(x,Q^2) + \Delta \bar q(x,Q^2)]$.
Similarly $\Delta q(x,Q^2)$'s (same for antiquarks) are defined as $\Delta q=q_+ - q_-$.
The gluon distributions are also defined as $G =G_+ + G_-$ and $\Delta G = G_+ - G_-$, but
in DIS they are not accessible directly and only enter in the QCD $Q^2$ evolution of the quark distributions.\\
There is a long list of interesting topics, {\it e.g.} characteristic features of unpolarized and polarized parton distributions, flavor separation of $\Delta q$, $\Delta \bar q$,
gluon polarization in the nucleon, generalized parton distributions,
quark transversity $\delta q(x,Q^2)$ and double transverse spin asymmetries $A_{TT}$,
single spin asymmetries (SSA) $A_N$ and QCD mechanisms, etc...\\
In this opening lecture, for lack
of time, we will have to make a strong selection, but given the high density of the scientific program, it will certainly allow to cover all missing important subjects. 

\newpage
\section{Digression on parton distributions functions}
 A new set of parton distribution functions (PDF) was constructed in the framework of a statistical approach of the nucleon \cite{bbs1}, which has the following characteristic
features:\\
 - For quarks (antiquarks), the building blocks are the helicity dependent distributions $q_{\pm}$ ($\bar q_{\pm}$) and we define $q= q_+ +q_-$ and $\Delta q = q_+ -q_-$ (similarly
 for antiquarks).\\
 - At the initial energy scale taken at $Q^2_0= 4 \mbox{GeV}^2$, these distributions are given
by the sum of two terms, a quasi Fermi-Dirac function and a helicity independent diffractive
contribution, which leads to a universal behavior for all flavors at very low $x$.\\
 - The flavor asymmetry for the light sea, {\it i.e.} $\bar d > \bar u$, observed in the data
is built in. This is clearly understood in terms of the Pauli exclusion principle, based on the fact that the proton contains two $u$ quarks and only one $d$ quark.\\
 - The chiral properties of QCD lead to strong relations between $q$ and $\bar q$.\\
For example, it is found that the well estalished result $\Delta u >0 $\ implies $\Delta 
\bar u >0$ and similarly $\Delta d <0$ leads to $\Delta \bar d <0$.\\
 - Concerning the gluon, the unpolarized gluon distribution is given in terms of a quasi
 Bose-Einstein function, with {\it no free parameter}, but for simplicity, one assumes zero gluon polarization, {\it i.e.} $\Delta G(x,Q_0^2)=0$, at the initial energy scale.\\
 - All unpolarized and polarized  distributions depend upon {\it eight}
free parameters, which were determined in 2002 (See \cite{bbs1}), from an NLO fit of a selected set of accurate DIS data.\\
For illustration, the $\pm$ light quark (antiquark) distributions are displayed on Fig. 1 and
we clearly notice the essential features mentioned above \footnote{For a practical use of these
PDF, see www.cpt.univ-mrs.fr/~bourrely/research/bbs-dir/bbs.html.}.
More recently, new tests against experimental (unpolarized and polarized)
data turned out to be very satisfactory, in particular in hadronic reactions \cite{bbs2,bbs3}.\\
\begin{figure}[thp]
\begin{center}
\includegraphics[width=65mm]{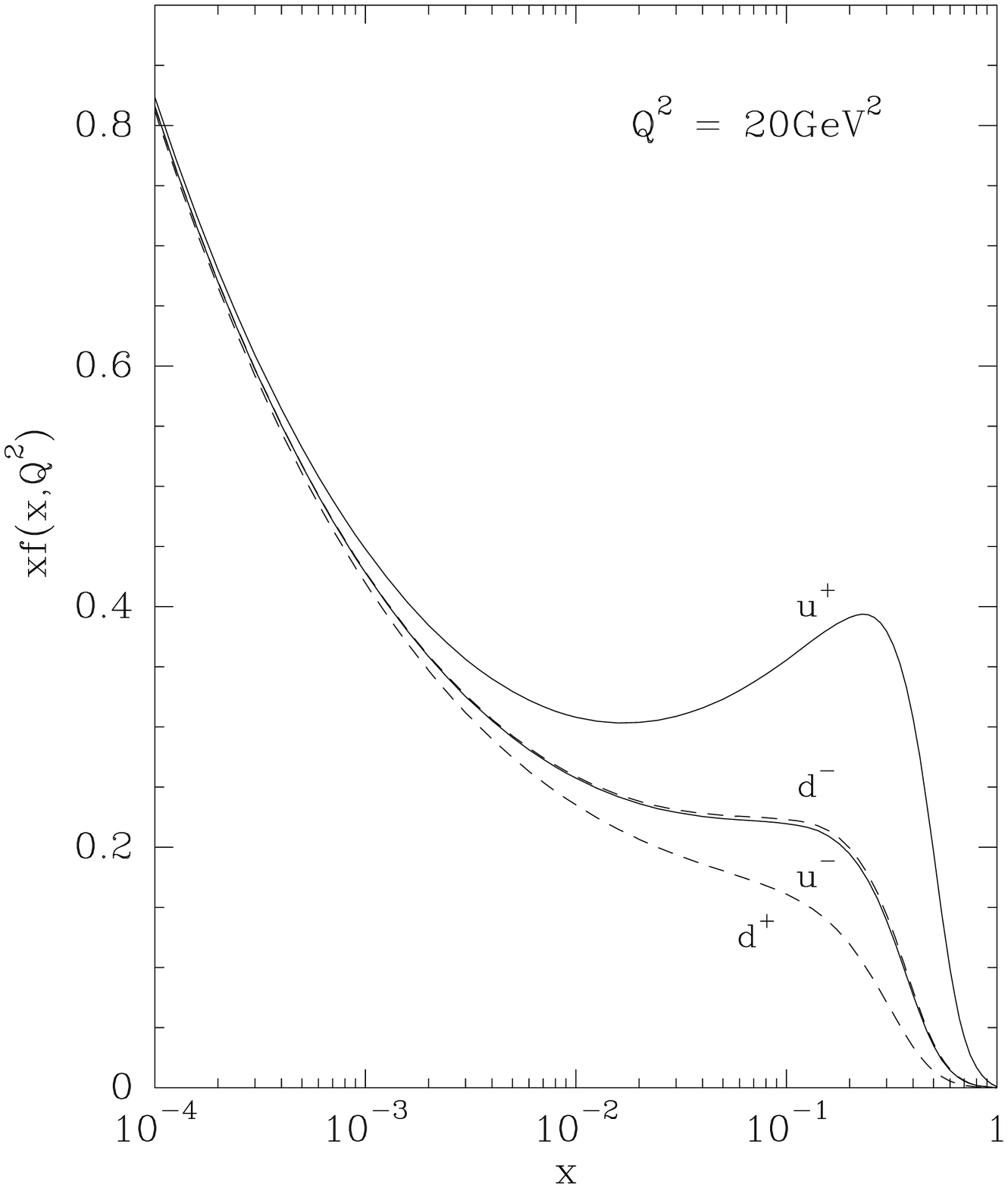}
\includegraphics[width=65mm]{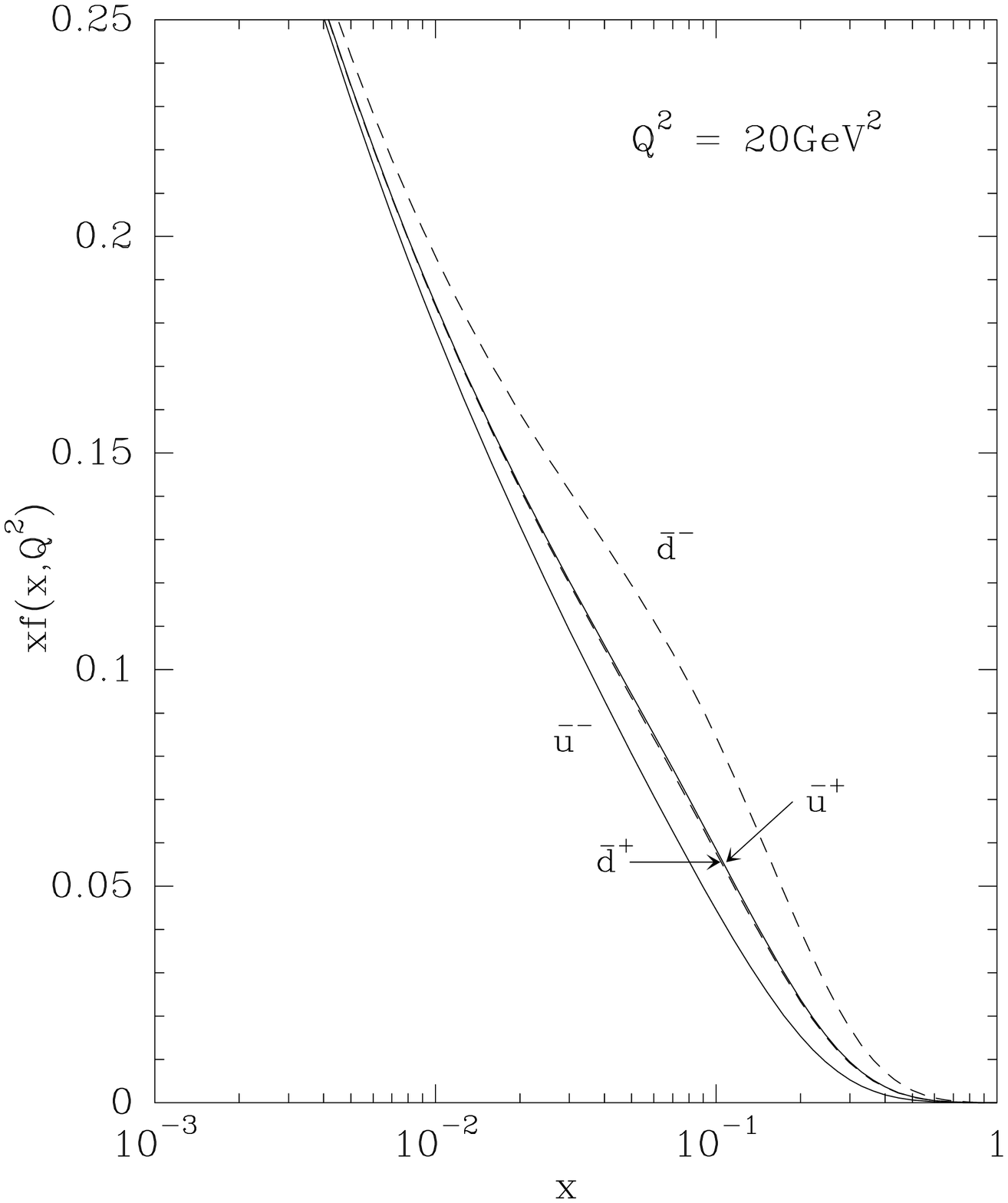}
\caption{\label{fig:1} On the left (right) the light quark (antiquark) distributions
with different helicities versus $x$ for $Q^2=20 \mbox{GeV}^2$, taken from Ref.~\cite{bbs1}.}
\end{center}
\end{figure}

The statistical approach has been extended to the interesting case where the PDF have, in addition to the usual Bjorken $x$ dependence, an explicit
transverse momentum $k_T$ dependence \cite{bbs4} and this might be used in future calculations
with no $k_T$ integration.
 Concerning the strange quark and antiquark distributions, a simplifying assumption consists to
take $s(x,Q^2)=\bar {s}(x,Q^2)$ and similarly for the corresponding polarized distributions
$\Delta s(x,Q^2)=\Delta \bar {s}(x,Q^2)$. However a careful analysis of the data led us to the conclusion that $s(x,Q^2)\neq \bar {s}(x,Q^2)$
and the corresponding polarized distributions are unequal, small and negative \cite{bbs5}.\\
Now let us come back to the important prediction of the statistical approach, namely
$\Delta \bar u >0$ and $\Delta \bar d <0$, which contrasts with the flavor symmetric assumption
$\Delta \bar u = \Delta \bar d = \Delta s = \Delta \bar s$ made, for example, in Ref.~\cite{hks}. With this assumption, the $\Delta \bar q$ don't contribute to the Bjorken sum rule, so one has to increase the absolute values of the valence contributions to $\Delta u$ and $\Delta d$, in order to satisfy
this sum rule. As shown on Fig.~2, this leads to over estimate $2xg_1^{(p-n)}(x)$ 
in the valence region, but it is not the case for the statistical approach. This has been confirmed by recent Compass data \cite{alek}.\\
\begin{figure}[thp]
\begin{center}
\includegraphics[width=8cm]{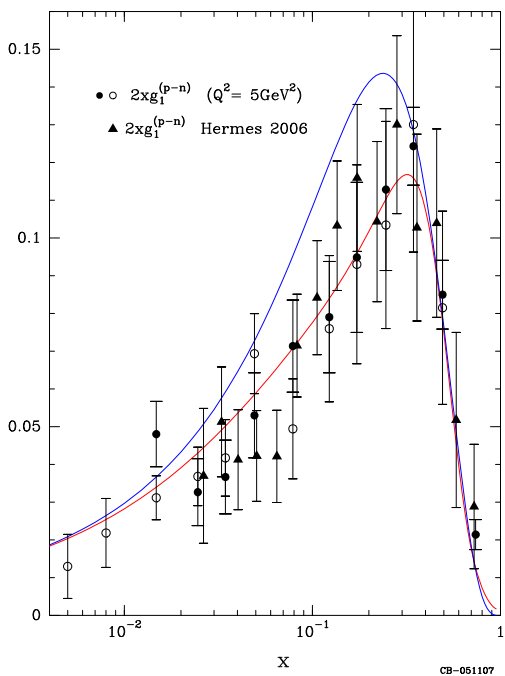}
\caption{\label{fig:2} Various recent data on the isovector structure function $2xg_1^{(p-n)}(x)$ compared to the statistical model prediction Ref.~\cite{bbs3} (lower curve)
and the AAC calculation Ref.~\cite{hks} (upper curve)(Taken from Ref.~\cite{bbs6}).}
\end{center}
\end{figure}
\begin{figure}[thp]
\begin{center}
\includegraphics[width=65mm]{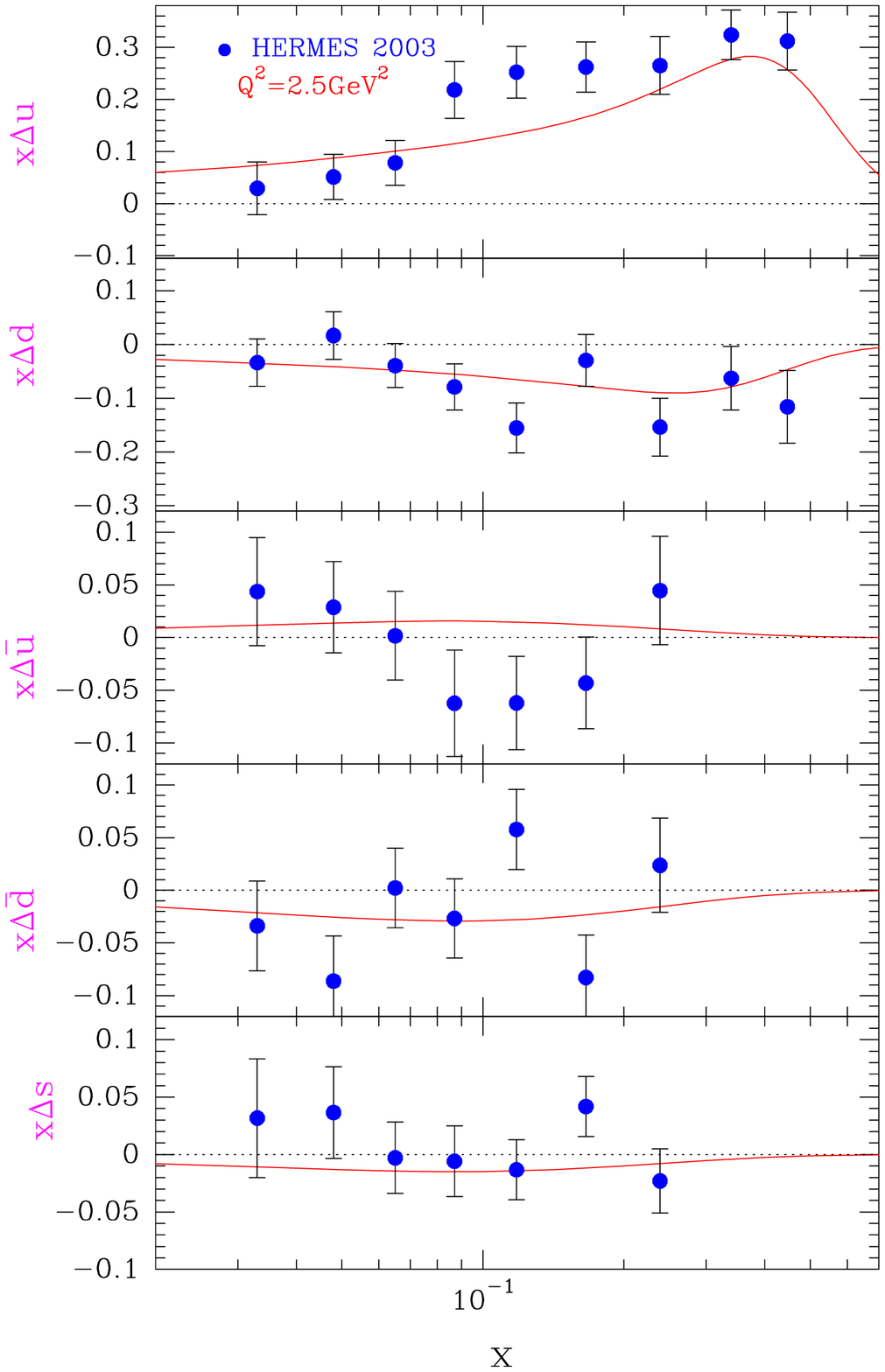}
\includegraphics[width=65mm]{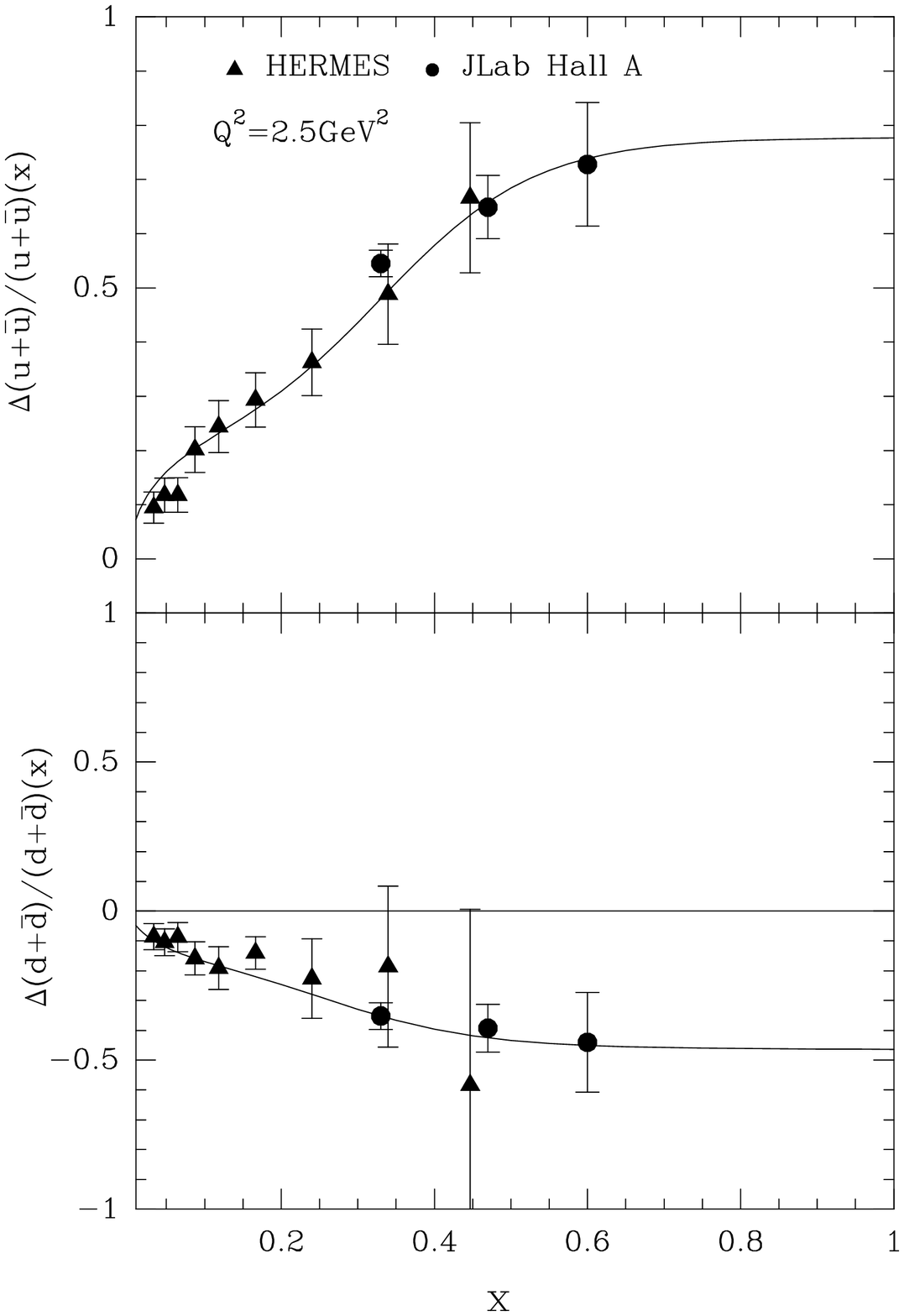}
\caption{\label{fig:3} Left: Quark and antiquark polarized parton distributions as a function of $x$ for $Q^2=2.5 \mbox{GeV}^2$. Data from Ref.~\cite{Hermes99}. Right: Ratios $(\Delta u +
\Delta \bar u)/(u + \bar u)$ and $(\Delta d + \Delta \bar d)/(d + \bar d)$ as a function of $x$.
Data from Hermes for $Q^2=2.5 \mbox{GeV}^2$ \cite{Hermes99} and a JLab experiment \cite{JLab}.
In both the curves are predictions from the statistical approach Ref.~\cite{bbs3}.}
\end{center}
\end{figure}
There is another way to test directly the predictions of the statistical approach for the polarized quark distributions and their flavor separation. This has been obtained from the
semi-inclusive polarized DIS and the Hermes data are shown on the left hand side of Fig.~3. On
the right hand side of Fig.~3, we also display the very accurate JLab data which show 
that, even in the high $x$ region, $\Delta (u+\bar u)$ remains positive whereas $\Delta (d+\bar d)$ remains negative, in accordance with the statistical approach expectations.\\
These features can and will be also investigated in future runs with polarized $pp$ collisions at BNL-RHIC, which we briefly discuss now.\\
Consider the parity-violating helicity asymmetry $A^{PV}_L(W)$
\begin{equation}
A_{L}^{PV}(y) = {\Delta d\sigma/dy \over d\sigma/dy}=
{ d\sigma^W_-/dy - d\sigma^W_+/dy \over d\sigma^W_-/dy + d\sigma^W_+/dy}~,
\end{equation}
where $\pm$ stands for the helicity of one polarized proton beam and $y$ is the $W$ rapidity. For $W^+$, at the lowest order of the Drell-Yan production mechanism, it reads 
\begin{equation}
A_{L}^{PV}(W^+) =
{\Delta u(x_a) \bar d(x_b) -\Delta \bar d(x_a)
u(x_b) \over u(x_a) \bar d(x_b)+ \bar d(x_a)
u(x_b)}~,
\end{equation}
where $x_a=\sqrt{\tau}e^y$, $x_b=\sqrt{\tau}e^{-y}$ and $\tau=M_W^2/s$.
For $W^-$ production one interchanges $u$ and $d$. 
The general trend of $A_{L}^{PV}(y)$ 
can be easily understood and, for example at $\sqrt s = 500\mbox{GeV}$  
near $y=+1$, $A_{L}^{PV}(W^+) \sim \Delta u /u$ and
$A_{L}^{PV}(W^-) \sim \Delta d /d$, evaluated at $x=0.435$. 
Similarly for near $y=-1$, $A_{L}^{PV}(W^+) \sim -\Delta \bar d /\bar d$ and
$A_{L}^{PV}(W^-) \sim -\Delta \bar u /\bar u$, evaluated at $x=0.059$.\\
The features appear clearly on the left hand side of Fig.~4, where the calculations
were done at two different energies. For completeness we also show the predicted $A_{L}^{PV}(Z)$
on the right hand side of Fig.~4, but in this case the interpretation is not so straightforward. Moreover the production rate of $Z$'s is much lower than $W$'s.\\ 

\clearpage
\newpage
\begin{figure}[thp]
\begin{center}
\includegraphics[width=65mm]{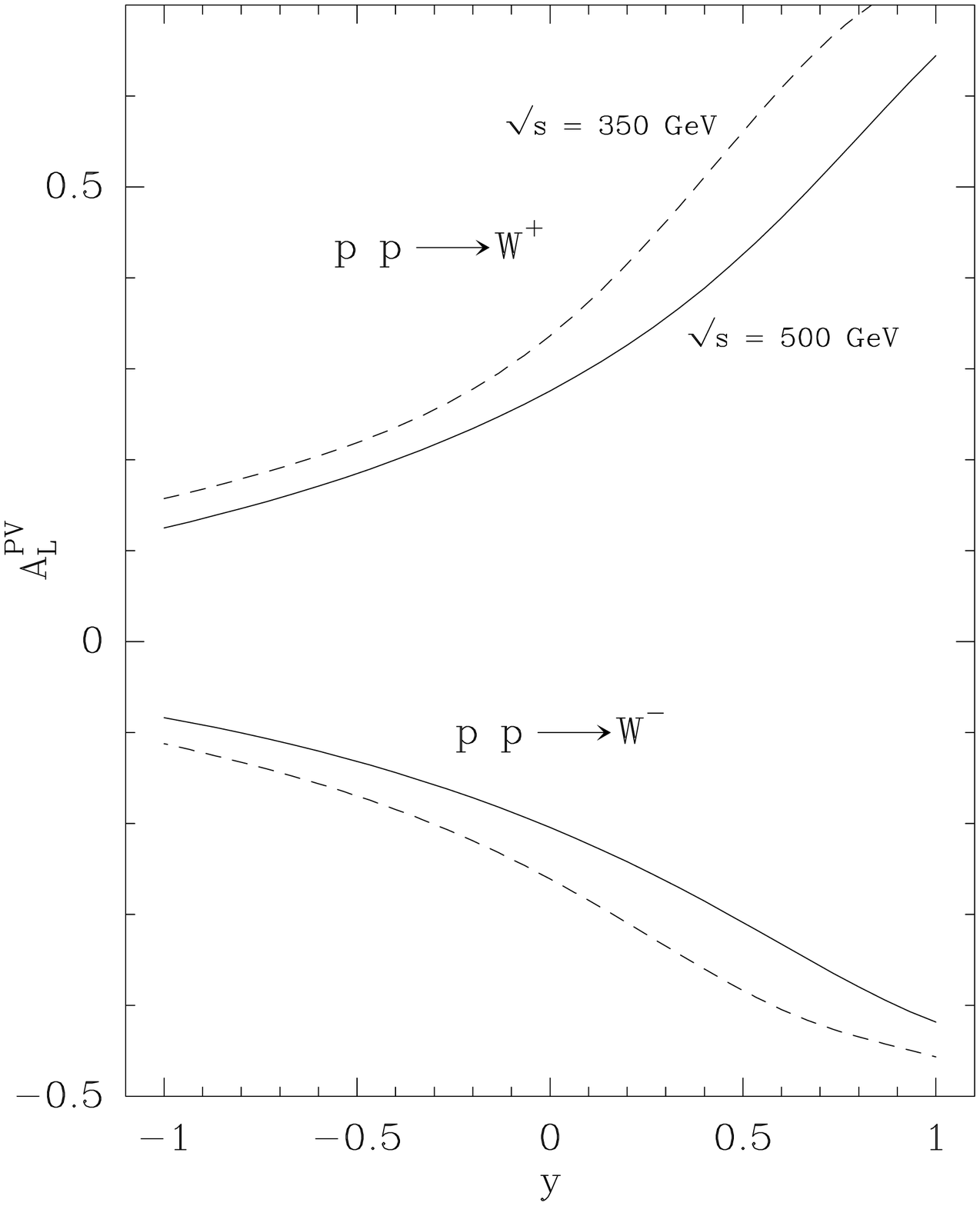}
\includegraphics[width=65mm]{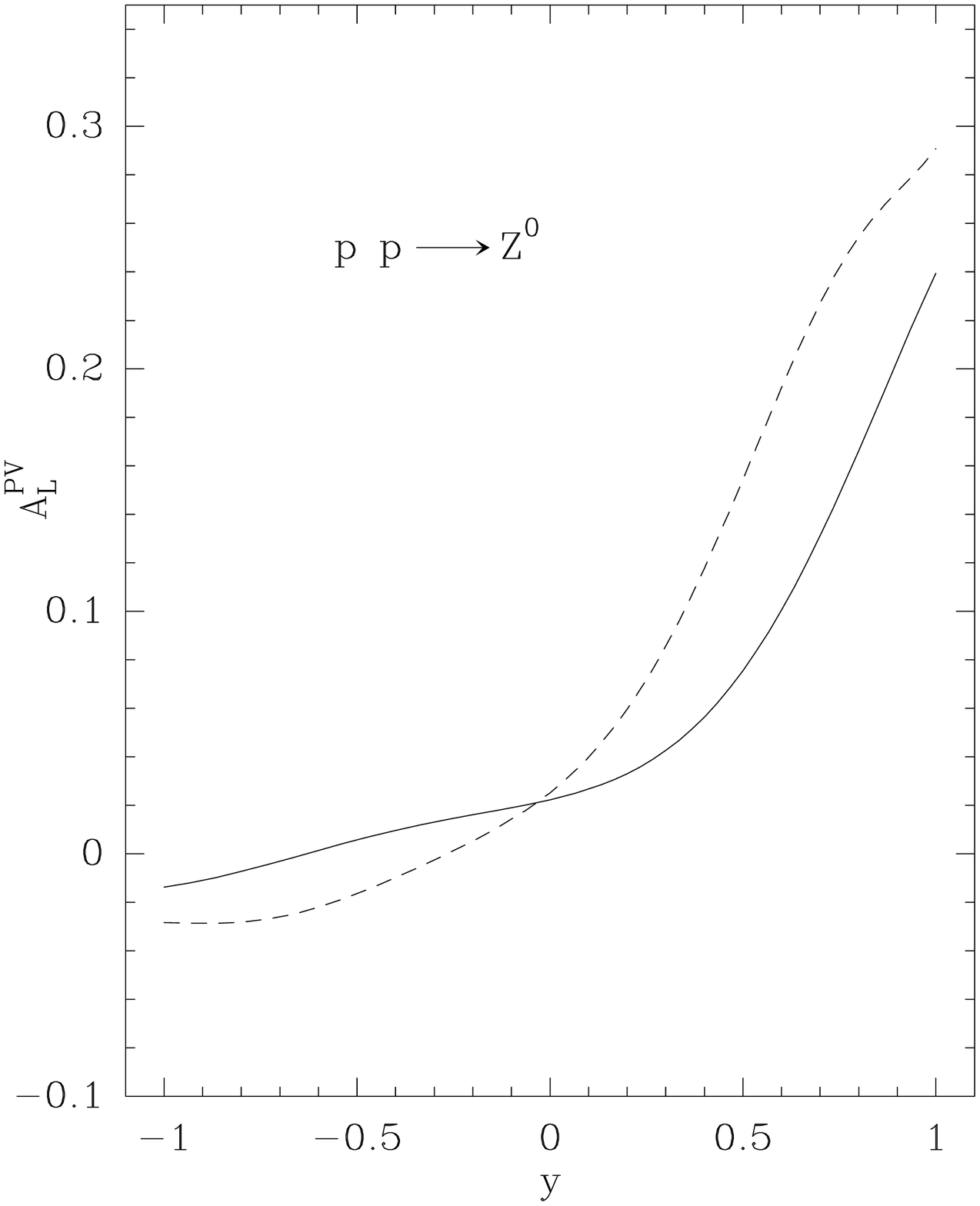}
\caption{\label{fig:4}Left: Predictions from the statistical appoach for the parity
violating asymmetry $A_L^{PV}$ for the $pp \to W^{\pm}$ production, versus the $W$ rapidity 
$y$, at
$\sqrt{s}=350$GeV (dashed curve) and $\sqrt{s}=500$GeV (solid curve). Right: Same for the
$pp \to Z$ production. (Taken from Ref.~\cite{bbs1})}
\end{center}
\end{figure}
However there is an important point to mention here, since the $W$'s are not directly seen.
For the most relevant signature, if one selects the leptonic decay $W \to e \nu$, one measures in fact
\begin{equation}
A_{L}^{PV}(y_e) = {\Delta d\sigma/dy_e \over d\sigma/dy_e}=
{ d\sigma^W_-/dy_e - d\sigma^W_+/dy_e \over d\sigma^W_-/dy_e + d\sigma^W_+/dy_e}~,
\end{equation}
where $y_e$ is the charged lepton rapidity. Fortunately, by using the RhicBos code due to P. Nadolski, one finds that $A_{L}^{PV}(y_e)$ has essentially the same trend as $A_{L}^{PV}(y)$.\\

So much for the quarks, let us now turn to the gluon distributions and we first consider
the unpolarized distribution $G(x,Q^2)$. In the statistical approach it has a very simple
expression (See Ref.~\cite{bbs1}), which is consistent with the available data, most coming indirectly from the QCD $Q^2$ evolution of $F_2(x,Q^2)$, defined earlier, in particular in the low $x$ region. However it is
known that $ep$ DIS cross section is characterised by two independent structure funtions, $F_2(x,Q^2)$ and the longitudinal structure function $F_L(x,Q^2)$. For low
$Q^2$, the contribution of the later to the cross section at HERA is only sizeable at $x$ smaller than approximately $10^{-3}$ and in this domain the gluon density dominates over the sea quark density. More precisely, it was shown that using some approximations, one has \cite{amcs}
\begin{equation}
xG(x,Q^2) = \frac{3}{10}5.9[\frac{3\pi}{2\alpha_s}F_L(0.4x,Q^2)-F_2(0.8x,Q^2)] \simeq \frac{8.3}{\alpha_s}F_L(0.4x,Q^2)~.
\end{equation}
Before HERA was shut down, a dedicated run period with reduced proton beam energy was
approved and we are waiting for these new H1 results on $F_L$. We show on Fig.~5 the predictions
of the statistical approach and the new data, whose precision is expected to be rather good, will allow to test its predictive power, once more.\\
\clearpage
\begin{figure}[thp]
\begin{center}
\includegraphics[width=8.0cm]{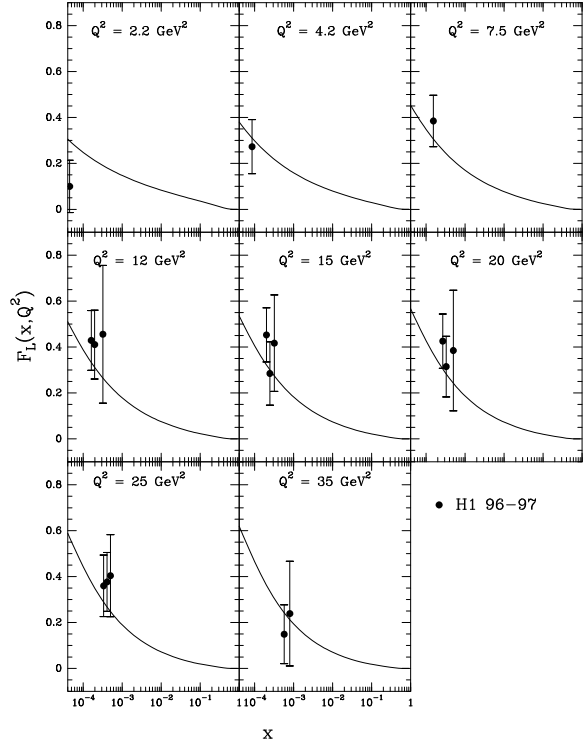}
\caption{\label{fig:5}Statistical approach predictions for the longitudinal
structure function $F_L(x,Q^2)$ with earlier H1 data (Taken from Ref.~\cite{bbs6}).}
\end{center}
\end{figure}
The polarized gluon distribution $\Delta G$ is also extremely important to determine and we
have the following helicity sum rule
\begin{equation}
\frac{1}{2} = \frac{1}{2}\Delta \Sigma + \Delta G(Q^2) + L_q(Q^2) + L_G(Q^2)~,
\end{equation}
where $\Delta \Sigma = \sum_q\int_0^1[\Delta q(x,Q^2) + \Delta \bar q(x,Q^2)]dx$
is twice the quark (+ antiquark) contribution to the nucleon helicity and
$\Delta G$, $L_{q,G}$ are the contributions of gluon and orbital angular momentum of
quark and gluon. So far $\Delta\Sigma \sim 0.3$ and the sum rule is not satisfied.\\
There are several attempts to extract $\Delta G(Q^2)$ from DIS using different processes
and the most recent results will be presented later in this Workshop. The RHIC spin program
is also putting a high priority to this determination and the cleanest reaction is inclusive prompt
photon production, which is dominated by the subprocess $G q \to \gamma q$. The double helicity
asymmetry, which has schematically the following expression
\begin{equation}
A_{LL} \approx \frac{\Delta G(x_1)}{G(x_1)} \cdot  
\Bigg[\frac{\sum_q e_q^2 \left[ \Delta q(x_2) + 
\Delta \bar{q}(x_2) \right]}{\sum_q e_q^2 \left[q(x_2)+
\bar{q}(x_2) \right] } \Bigg] \cdot 
\hat{a}_{LL}(Gq \rightarrow \gamma q) + (1\leftrightarrow 2)~, 
\end{equation}
is directly proportional to $\Delta G$. This has not been measured yet, but from the measurements on $\overrightarrow{p} \overrightarrow{p} \to \pi(or jet) X$, we have all indications that $\Delta G$ is small and still badly known, unfortunately.\\
The next very serious question is indeed: are there relevant contributions from $L_{q,G}$?  
\clearpage

\section{Quark transversity $\delta q(x,Q^2)$ and $A_{TT}$}
The existence of this new quark distribution $\delta q(x,Q^2)$,
was first mentioned by Ralston and Soper in 1979, by studying the angular distribution in $p(\uparrow)p(\uparrow) \to \mu^+\mu^- X$ with
transversely polarized protons. It was merely forgotten until 1990, where it was first realized that it
completes the description of the quark distribution in a nucleon as a density matrix
\begin{equation}
{\cal Q}(x,Q^2)=q(x,Q^2)I\otimes I + \Delta q(x,Q^2)\sigma_3\otimes\sigma_3
+ \delta q(x,Q^2)(\sigma_+\otimes\sigma_- + \sigma_-\otimes\sigma_+)~.
\end{equation}
This quark transversity $\delta q(x,Q^2)$ is chiral odd, leading twist and
decouples from DIS. So it was never measured and we only have
the following positivity bound \cite{js} \footnote{Positivity is extremely usefull to constrain spin observables, as discussed by X. Artru in these proceedings (See also Ref.~\cite{aerst}).} 
\begin{equation}
q(x,Q^2) + \Delta q(x,Q^2) \geq 2|\delta q(x,Q^2)|~, 
\end{equation}
which survives up to NLO corrections. It is indeed accessible in $p(\uparrow)p(\uparrow) \to \mu^+\mu^- X$, with both protons transversely polarized. The double transverse spin asymmetry 
$A_{TT}$ reads
\begin{equation}
A_{TT} = \frac{d\sigma(\uparrow\uparrow) - d\sigma(\uparrow\downarrow)}
{d\sigma(\uparrow\uparrow) + d\sigma(\uparrow\downarrow)} = \widehat a_{TT} \frac{\sum_{q}^{} e^2_q \delta q
(x_1,M^2) \delta \bar{q} (x_2,M^2) + (1 \leftrightarrow 2)}
{\sum_{q}^{} e^2_q  q (x_1,M^2) \bar q (x_2,M^2) + 
(1 \leftrightarrow 2)}~,
\end{equation}
where $\widehat a_{TT}=-1$ and $M^2$ is the dilepton mass square. It
involves the product of $\delta q$ and $\delta \bar q$, as expected from the dominant $q \bar q$
annihilation Drell-Yan mechanism. Predictions using
the saturation of the bound, lead to some estimates of only a few percents, but it is on the
list of future measurements at the BNL-RHIC spin program.\\
The asymmetry at the $Z$ pole, which reads
\begin{equation}
A_{TT}(Z) = \frac{\sum_{q}^{} (b^2_q - a^2_q) \delta q
(x_1,M_Z^2) \delta \bar{q} (x_2,M_Z^2) + (1 \leftrightarrow 2)}
{\sum_{q}^{} (b^2_q + a^2_q) q (x_1,M_Z^2) \bar q (x_2,M_Z^2) + 
(1 \leftrightarrow 2)}~,
\end{equation}
is also expected to be small. However, for the $W^{\pm}$ production, considered above, $A_{TT}=0$, because the $W$ has a $V-A$ coupling, {\it i.e.} $a_q = b_q$, which remains to be checked.\\
There is no such a transversity distribution for gluons which carry a spin one and this fact has
important consequences for $A_{TT}$ of different reactions. For example in the case of single-jet production, according to pQCD, the cross section in the low $p_T$ region is
dominated by gluon- gluon collisions, in the medium $p_T$ region by gluon-quark collisions
and in the high $p_T$ region by quark-quark collisions. As a result, $A_{TT}$ is expected to
be non-zero only in this last kinematic region and this is what we see on the left hand side of
Fig.~6. We have a similar situation for prompt photon production, shown on the right hand side
of Fig.~6. These results, which were obtained by using the positivity bound, probe the sensitivity only to quark transversity in the hight $p_T$ region. As was noticed in Ref.~\cite{ssv}, we expect double spin transverse asymmetries to be much smaller than double
helicity asymmetries, {\it i.e.} $|A_{TT}|<<|A_{LL}|$ and this theoretical observation must be
carefully confirmed experimentally.
\clearpage
\begin{figure}[thp]
\begin{center}
\includegraphics[width=79mm]{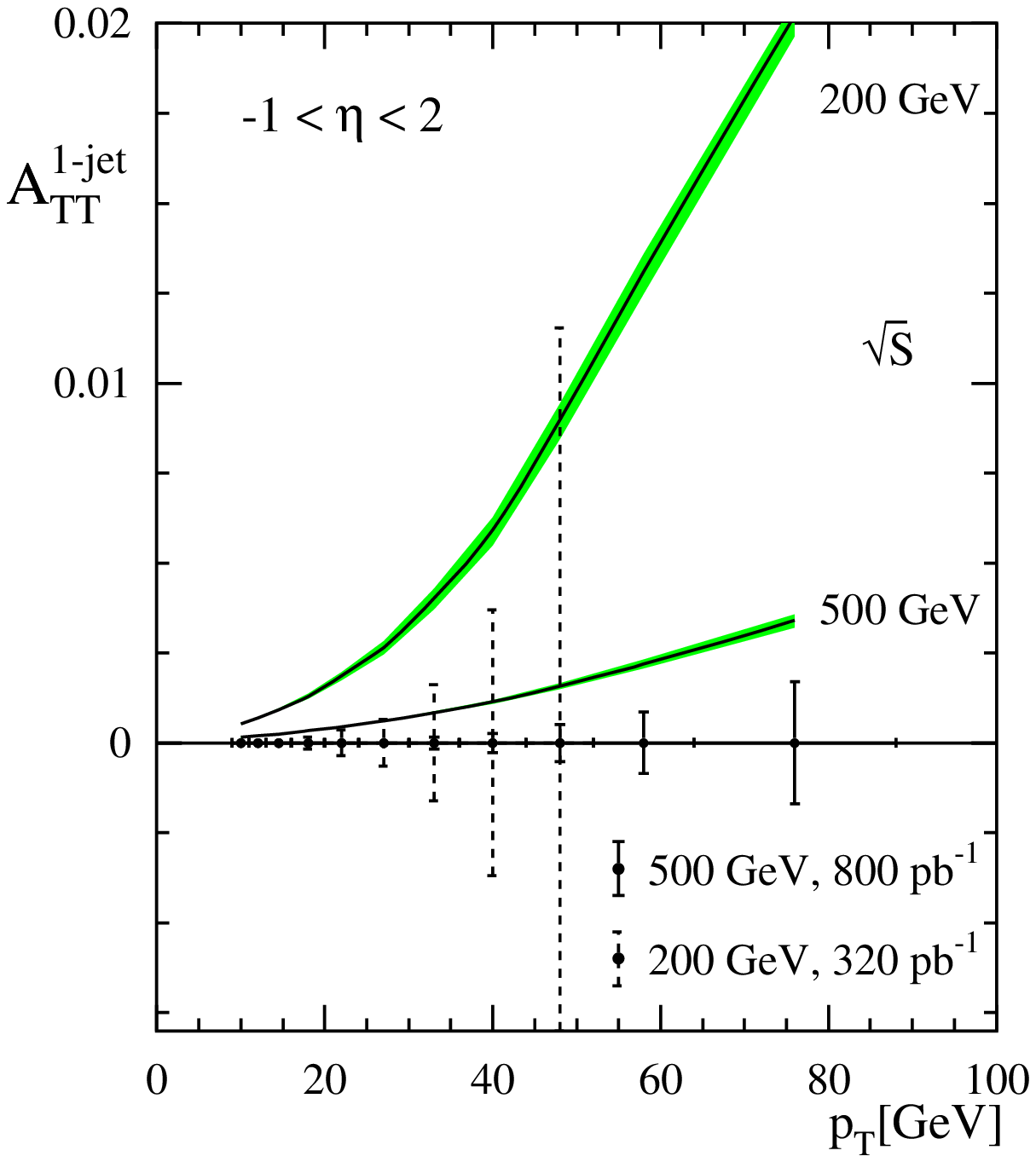}
\includegraphics[width=79mm]{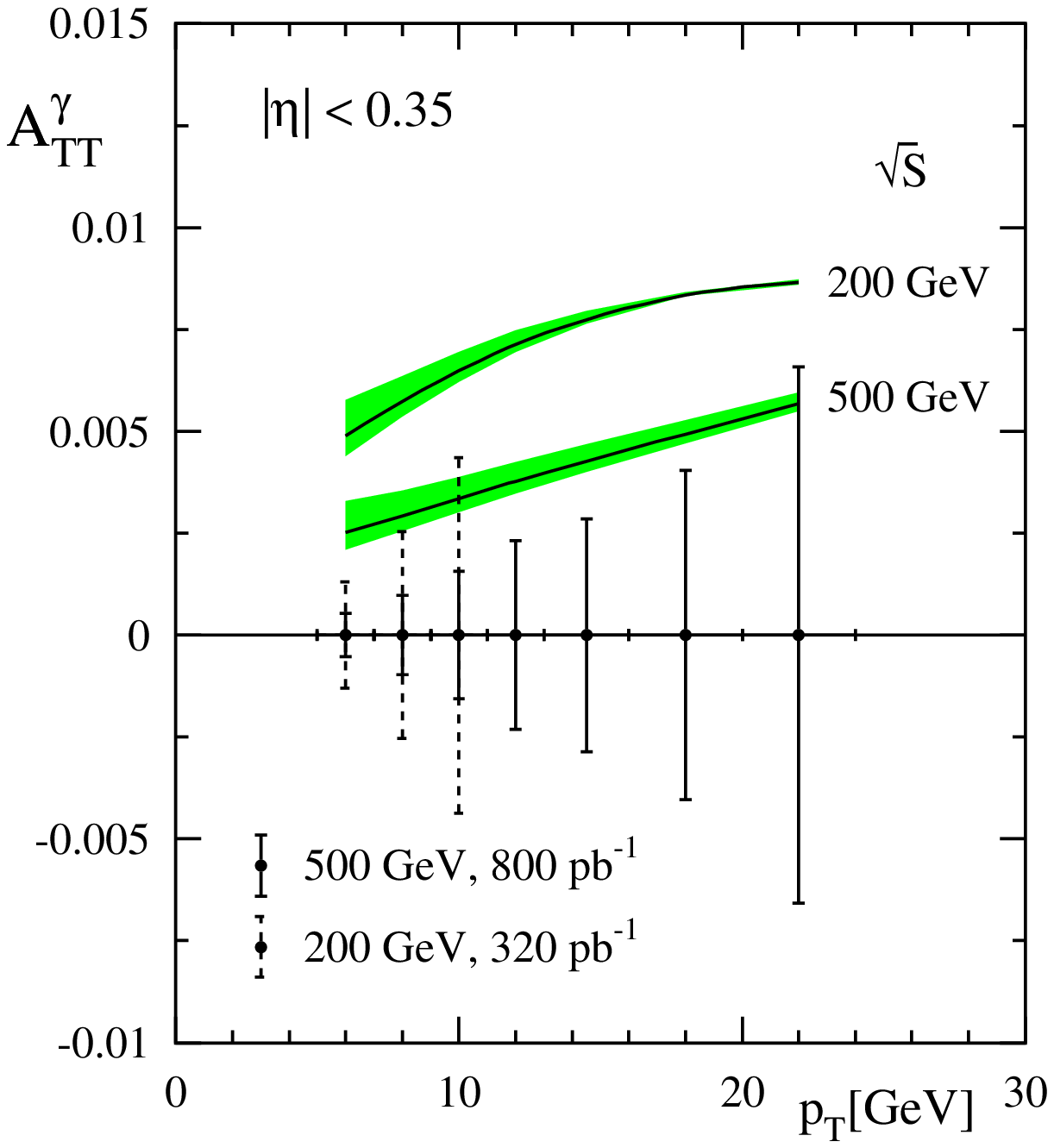}
\caption{\label{fig:6}Left: Upper bounds for $A_{TT}$ for single jet
production at RHIC, with the expected statistical errors. Right: same for
prompt photon production (Taken from Ref.~\cite{ssv}).}
\end{center}
\end{figure}
\section{Single spin asymmetry in QCD}
 What is a single spin asymmetry (SSA)?\\
Consider the collision of a proton of momentum $\overrightarrow{p}$, carrying
a transverse spin $\overrightarrow{s_T}$ and producing an outgoing hadron with
transverse momentum $\overrightarrow{k_T}$. The SSA defined as
\begin{equation}
A_N = \frac{d\sigma(\overrightarrow{s_T}) - d\sigma(-\overrightarrow{s_T})}{d\sigma(\overrightarrow{s_T}) + d\sigma(-\overrightarrow{s_T})}
\end{equation}
is zero, unless the cross section contains a term $\overrightarrow{s_T}\cdot(\overrightarrow{p}\times \overrightarrow{k_T})$.
It can be shown that this requires the existence of an {\it helicity flip}
and {\it final state interactions}, which generate a phase difference between 
the flip and the non-flip amplitudes, to avoid violation of time reversal invariance.
In the naive parton model one expects very small SSA, because of the double
suppression $\alpha_s m_q/Q$, where $m_q$ is the quark mass and $Q$ the energy scale of the process.\\
Actually a large SSA has been discovered 30 years ago 
at FNAL with a 300 GeV/c unpolarized proton beam in $pBe \to \Lambda^{\uparrow} X$ \cite{bunce} and many more SSA have been observed later, in particular large
SSA in $p^{\uparrow}p \to \pi X$ and $\bar {p}^{\uparrow}p \to \pi X$ at FNAL by E704 \cite{E704} and more recently by STAR at BNL-RHIC \cite{STAR}. These data have the same trend,  as shown in Fig.~7, although they were obtained in very different energy ranges. Therefore one
can be tempted to conclude that they originate from the same mechanism satisfying scaling.\\
Before discussing this point, we recall that in the collinear approximation, the mechanism to generata SSA is based on higher-twist quark-gluon correlators (Efremov-Teryaev 1982, Qiu-Sterman 1991). However, if one
introduces transverse momentum dependence (TMD), two QCD mechanisms have been proposed:\\
 - TMD parton distributions $\Rightarrow$ Sivers effect 1990\\
 - TMD fragmentation distributions $\Rightarrow$ Collins effect 1993\\
The gauge-invariance properties of the TMD PDF have been first clarified for DIS and Drell-Yan processes in Ref.~\cite{bhs}.
In general both Sivers and Collins effects contribute to
a specific reaction, although there are some cases in which
only one of them contributes. For example in semi inclusive DIS, the Collins effect is
the only mechanism that can lead to asymmetries $A_{UT}$ and
$A_{UL}$. On the other hand, it does not appear in
some electroweak interaction processes, where there is only the
Sivers effect. In prompt photon production in $pp$ collisions, which
is dominated by $qG \to q \gamma$, the SSA is sensitive to either the
quark or the gluon Sivers functions, according to the value of the photon $x_F$ \cite{ssy}.\\
Now let us ask:
do we understand the SSA displayed on Fig.~7, given the fact that STAR is at a very small
angle 2.6 deg., whereas E704 is at a much larger angle, between 9 deg. and 64 deg.? A negative answer
is partially obtained by looking at the cross section.
The pQCD NLO calculation underestimates the cross section at {\it low}
energies and {\it medium} angles, namely for the E704 kinematic region. This is shown on Fig.~7
and it means that one should not ignore other contributions. This is not the case at 90 deg. and at very small
angles at high energy, which is the STAR kinematic range. To conclude, one should not
try to "explain" the SSA, ignoring the unpolarized 
cross section \cite{BS}. Of course one should not forget
resummation effects, which might help clarifying the situation.
\begin{figure}[thp]
\begin{center}
\includegraphics[width=70mm]{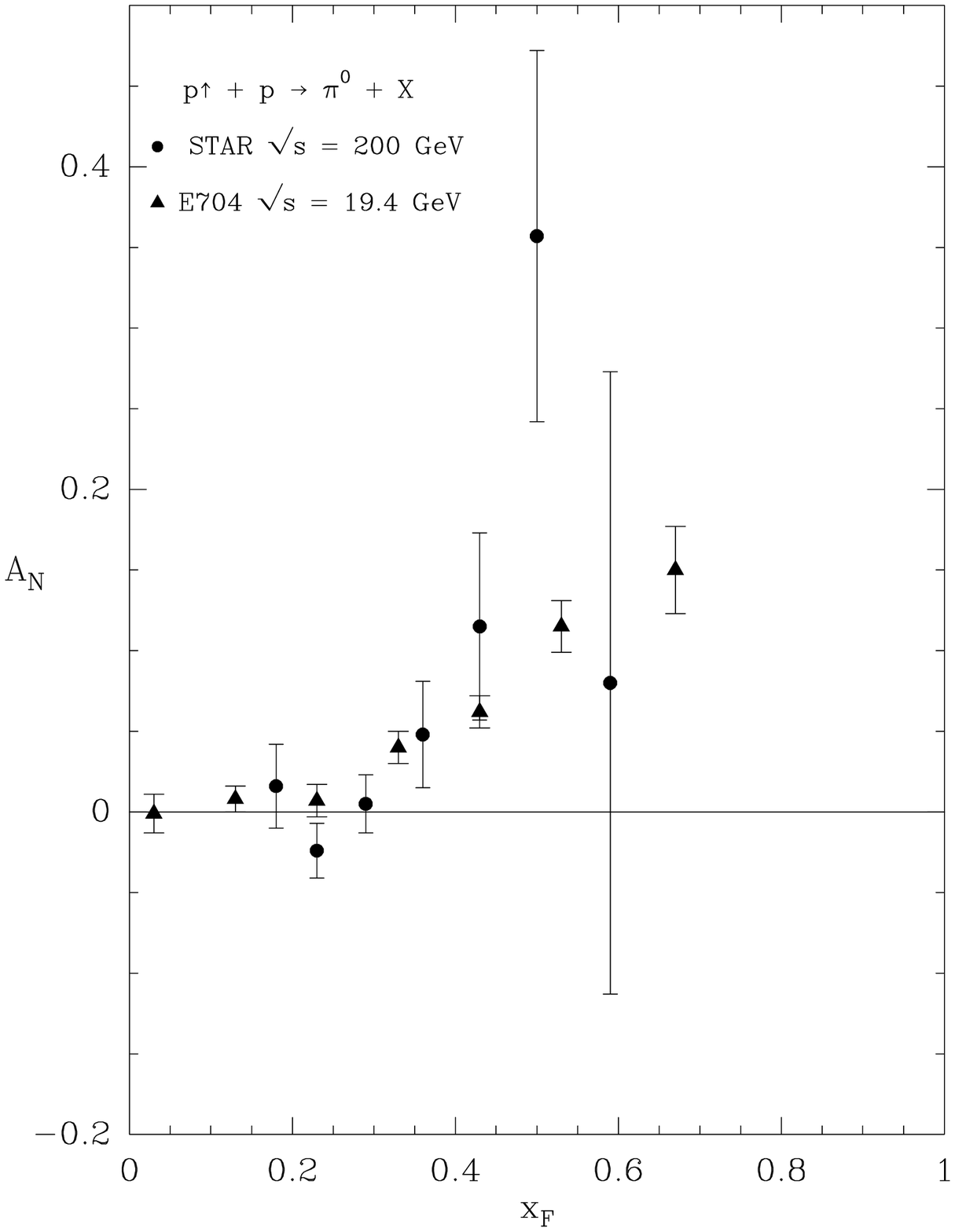}
\includegraphics[width=70mm]{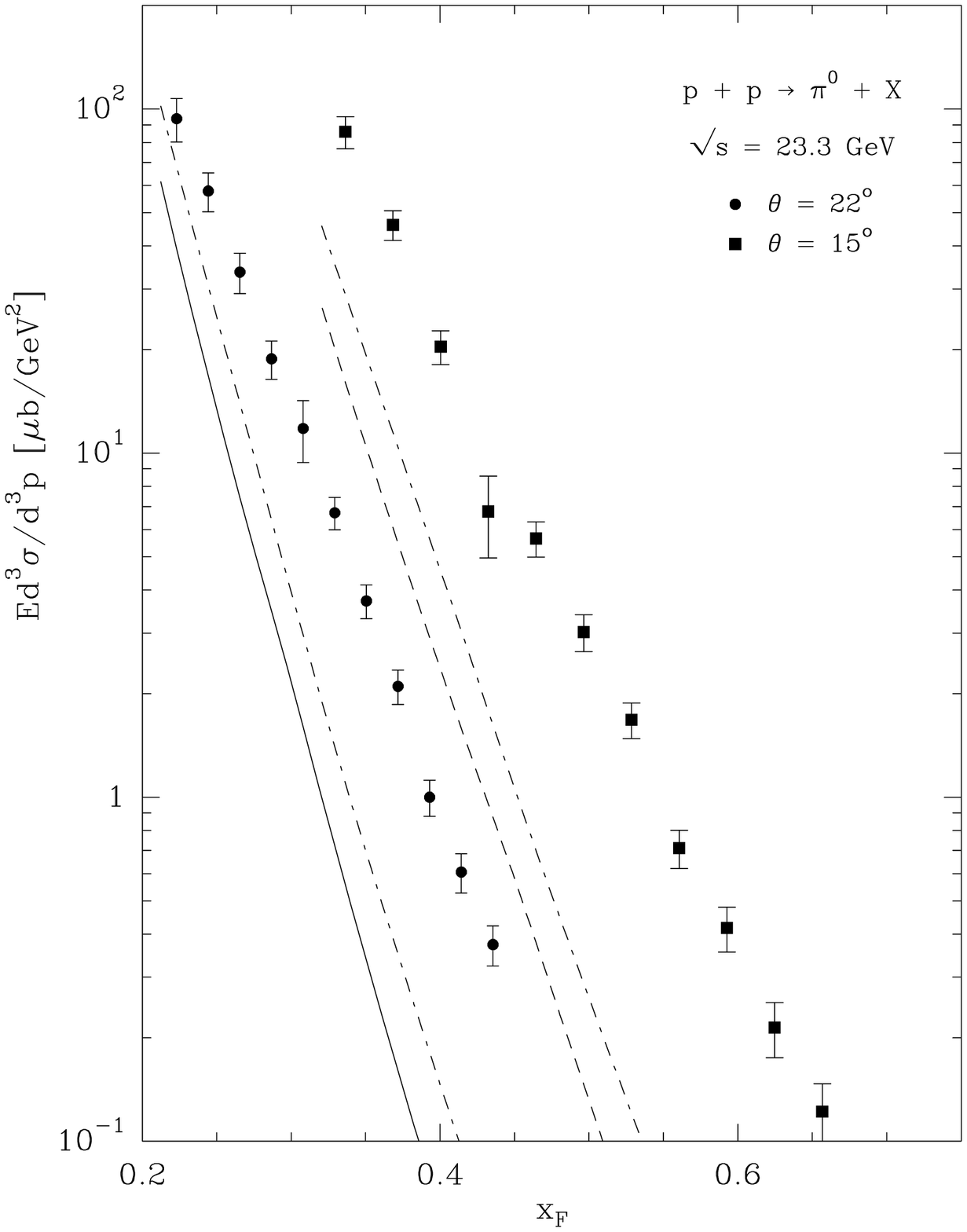}
\caption{\label{fig:7}Left: The single spin asymmetry $A_N$ as a function of $x_F$
at two different energies. The data are from Refs.~\cite{E704,STAR}. Right: A comparison
between a pQCD NLO calculation and data for two different angles ( Taken from Ref.~\cite{BS}) .}
\end{center}
\end{figure}
\clearpage
{\bf Acknowledgements}\\
This work was completed during my visit at the Universidad Santa Maria, Valparaiso, supported by the cooperation program Ecos-Conicyt C04E04 between France and Chile.\\
I am grateful to the organizers of DSPIN07, for their invitation to this conference dedicated to L. I. Lapidus, I had the great privilege to meet several times. My special thanks go also to Prof. A.V. Efremov for providing a full financial support and for making, once more, this meeting so successful.\\  

\end{document}